\documentclass[reprint, twocolumn, amssymb, amsmath, bibnotes, aps, prl, superscriptaddress,longbibliography]{revtex4-2}

\usepackage{graphicx,comment}
\usepackage{float}
\usepackage{dcolumn}
\usepackage{bm}
\usepackage{hyperref}
\usepackage[dvipsnames]{xcolor}
\hypersetup{
    colorlinks=true,       
    linkcolor=blue,     
    citecolor=blue,    
    urlcolor=blue      
}
\usepackage{natbib}
\usepackage[caption=false]{subfig}
\usepackage{color,soul}
\usepackage{environ}
\usepackage{upgreek}
\usepackage{pdfpages} 
\usepackage{pgffor} 

\makeatletter
\AtBeginDocument{\let\LS@rot\@undefined}
\makeatother

\def\supplementfilename{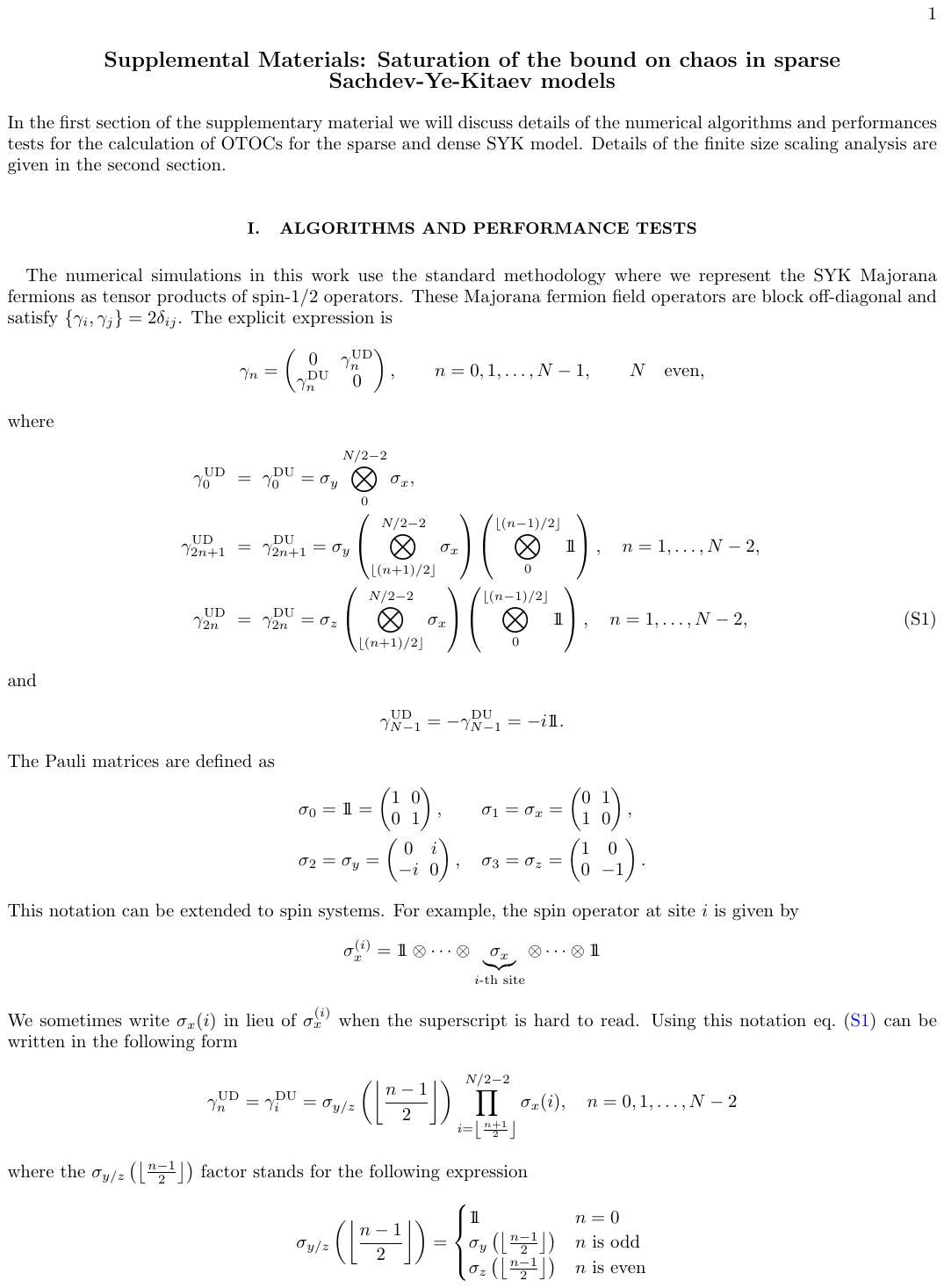}

\pdfximage{\supplementfilename}
\def\numbersupplementpages{\the\pdflastximagepages}

\newif\ifarXiv
\arXivtrue

\newcommand{\del}{\partial}

\newcommand{\eref}[1]{(\ref{#1})}
\newcommand{\nn}{\nonumber}
\newcommand{\be}{\begin{eqnarray}}
\newcommand{\ee}{\end{eqnarray}}

\newcommand{\bmat}{\left ( \begin{array}{cc} }
\newcommand{\emat}{\end{array} \right ) }

\def\Tr{\textrm{Tr}}

\newcounter{amg}

\newcounter{cl}

\newcounter{jvc}

\begin{document}

\title{Sparsity independent Lyapunov exponent in the Sachdev-Ye-Kitaev model}
\author{Antonio M. Garc\'\i a-Garc\'\i a}
\email{amgg@sjtu.edu.cn}
\affiliation{Shanghai Center for Complex Physics,
	School of Physics and Astronomy, Shanghai Jiao Tong
	University, Shanghai 200240, China}
\author{Chang Liu}
\email{cl91tp@gmail.com}
\affiliation{Shanghai Center for Complex Physics,
	School of Physics and Astronomy, Shanghai Jiao Tong
	University, Shanghai 200240, China}
\author{Jacobus J. M. Verbaarschot}
\email{jacobus.verbaarschot@stonybrook.edu}
\affiliation{Center for Nuclear Theory and Department of Physics  Astronomy, Stony Brook University, Stony Brook, New York 11794, USA}

\begin{abstract}
  The saturation of a recently proposed universal bound on the Lyapunov exponent has been conjectured to signal the existence of a gravity dual. This saturation occurs in the low temperature limit of the dense Sachdev-Ye-Kitaev (SYK) model, $N$ Majorana fermions with $q$-body ($q>2$) infinite-range interactions.
   We calculate certain Out of Time Order
  Correlators (OTOC) for $N\le 64$ fermions for a highly sparse SYK model and find no significant dependence
  of
  the Lyapunov exponent on sparsity up to near the percolation limit where the Hamiltonian breaks up into blocks.
  This suggests that in the sparse case, the Lyapunov exponent also saturates the low-temperature bound.
  A key ingredient to reaching $N = 64$ is the development of a novel quantum spin model simulation library 
  that implements highly-optimized matrix-free Krylov subspace methods on Graphical Processing Units (GPUs). This leads to a significantly lower simulation time as well as vastly reduced memory usage over previous approaches, while using modest computational resources. Strong sparsity-driven statistical fluctuations require both the use of a vastly larger number of disorder realizations with respect to the dense limit 
  and a careful finite size scaling analysis. Our results potentially broadens the landscape of theories that may have a gravity analogue.

\end{abstract}

\maketitle

The exponential growth of certain out-of-time-order correlation functions (OTOC) up to  the Ehrenfest time in the semiclassical limit, at a rate given by the leading classical Lyapunov exponent,
is an early signature of quantum chaotic dynamics. Their
calculation in simple single-particle problems such as a particle in a random potential \cite{larkin1969} or kicked rotors \cite{berman1978} were landmarks in the early development of the theory of quantum chaos.
However, they are notoriously difficult to compute quantitatively in many-body systems because the region of exponential growth is relatively short and can be easily overshadowed by other contributions unless the system is strictly within the semiclassical limit.

A resurgence of interest in OTOCs in quantum chaos came quite unexpectedly
from quantum gravity. Heuristic arguments \cite{sekino2008,shenker2013}
suggest that the dynamics of a particle close to a black hole horizon is quantum chaotic. Later, these ideas were put on a much firmer ground by showing that a universal bound on the Lyapunov exponents in quantum chaotic systems at thermal equilibrium was saturated in field theories with a gravity dual \cite{maldacena2015}. Shortly afterwards, Kitaev \cite{kitaev2015} demonstrated analytically that, in the low temperature limit, this universal bound on chaos was saturated in a simple model, now termed the Sachdev-Ye-Kitaev model (SYK) \cite{kitaev2015,bohigas1971,french1970,bohigas1971,french1971,sachdev1993,benet2001}, consisting of $N$ Majoranas \cite{kitaev2015,maldacena2016} with random $q$-body interaction in zero dimensions. The quantum chaotic nature of the SYK model for longer time scales was confirmed by a level statistics analysis \cite{garcia2016,cotler2016} and its gravity dual was identified to be Jackiw-Teitelboim gravity \cite{jackiw1985,teitelboim1983,almheiri2015}.

The analytical tractability of the SYK model is one of its most appealing features.
 Unfortunately, generalizations of the model with finite range  \cite{garcia2019} or sparsified \cite{garcia2021,swingle2020,caceres2021,caceres2022,tezuka2023} interactions do not inherit this property. 
 This begs the question: is the saturation of the bound, that indicates the possible existence of a gravity dual, a particularity of the dense SYK, or is it present in more general settings?
 For the dense SYK, a recent numerical calculation \cite{kobrin2020} of the Lyapunov exponent based on the Krylov subspace method \cite{krylov1931,krylov2007}
for up to $N = 50$ on a GPU-system (and $N = 60$ on a CPU-only system) confirmed the analytical results \cite{kitaev2015,maldacena2016}.
An important benefit of the sparsified SYK model is that it may be easier to simulate
on a quantum computer \cite{Jafferis:2022crx,Kobrin:2023rzr,Jafferis:2023moh}
which potentially facilitates addressing questions
 that cannot be answered with classical computers.

In this paper, we aim to calculate the Lyapunov exponent for a sparse variant
\cite{garcia2021,swingle2020,caceres2021,caceres2022,tezuka2023} of the SYK model where a large (to be defined shortly) number of random couplings are set to zero. A key ingredient in our study is the development of a highly-optimized GPU computing code, which implements the Krylov-based algorithm for computing time evolution of qubit systems. This allows us to reach up to $N = 64$ Majoranas on single GPU systems.
Our main result is that  the Lyapunov exponent of the sparse SYK model has
no significant dependence on sparsity, and agrees
with the dense case, all the way up to close to the percolation limit. This suggests the existence of gravity duals in a much broader family of field theories.
\medbreak
{\noindent\it Sparse SYK Model.}\quad Our Hamiltonian describes $N$ strongly interacting Majorana fermions in zero
spatial dimensions \cite{kitaev2015,bohigas1971,bohigas1971a,french1970,mon1975,french1971,benet2003,sachdev1993,maldacena2016} with sparse \cite{garcia2021,swingle2020,caceres2021,caceres2022,caceres2023} random interactions of infinite range:
\begin{equation}\label{hami}
  H = \sum_{0\leq i<j<k<l< N} p_{ijkl}\, J_{ijkl} \, \gamma_i \gamma_j \gamma_k \gamma_l \, .
\end{equation}
The Majorana  operators $\gamma_n$ satisfy the Clifford algebra
$\{\gamma_m , \gamma_n \}=\delta_{mn}$, and can be expressed as a tensor product
of Pauli matrices.
The $J_{ijkl}$ are random numbers with a Gaussian distribution of zero average and variance $\langle J^2 \rangle = {3}/({8pN^3})$.
The sparseness of the Hamiltonian is modeled by the stochastic
variable $p_{ijkl}$ which is sampled from the Bernoulli distribution $B(p)$ with probability $p>0$. When $p=1$,  we recover the dense SYK model. Models with $0<p<1$ are called sparse SYK models.
In principle, $p \sim N^{-\alpha}$ with  parameter $\alpha > 0$. It was shown in Ref.~\cite{swingle2020,garcia2021}, $\alpha = 3$ is the relevant scaling to study the effect of sparsity because for $\alpha > 3$ connectivity in Fock space is broken in the large $N$ limit while the effect of sparsity is largely irrelevant for $\alpha < 3$. Therefore, it is natural to define the sparsity strength 
$k = \frac {p}N \binom{N}{4}$.
  For the comparison with the dense case, we will focus on
  the $ k \geq 3$ region only because, for sufficiently large $N$, it is computationally expensive to impose
  a regularity condition on the vertex connectivity. The latter is necessary for $k \sim 1$ in order to prevent the Hilbert space from splitting into separate invariant subspaces of the Hamiltonian.
\medbreak
{\noindent\it OTOC calculation and results.}\quad We now define the following regularized out-of-time-order correlation (OTOC) function for the Hamiltonian Eq.(\ref{hami}),
 \be\label{eq:otoc}
 F(t)&=&\frac 1Z \Tr \left[e^{itH} \gamma_{\scriptscriptstyle N-1} e^{-(it + \beta /4)H} \gamma_{\scriptscriptstyle N-2}\right . \nn\\ &&\left . 
 e^{(it-\beta /4)H} \gamma_{\scriptscriptstyle N-1} e^{-(it +\beta /4)H} \gamma_{\scriptscriptstyle N-2} e^{-\beta H/4} \right ]
 \ee
with $Z= \Tr \left (  e^{-\beta H/4} \gamma_{\scriptscriptstyle N-1} e^{-\beta H /4} \gamma_{\scriptscriptstyle N-2} e^{-\beta H /4} \gamma_{\scriptscriptstyle N-1}e^{-\beta H /4}\right.$ $\left.\gamma_{\scriptscriptstyle N-2} \right)$ so that $F(0) =1$.
Different regularizations may lead to slightly different prefactors in the $1/N$ expansion of  the OTOC which may be time dependent but the Lyapunov exponent was recently shown \cite{tsuji2018bound,bermudez2019} to be independent of the regularization.

\begin{figure}[t!]
  \centering
  \includegraphics[width=6cm]{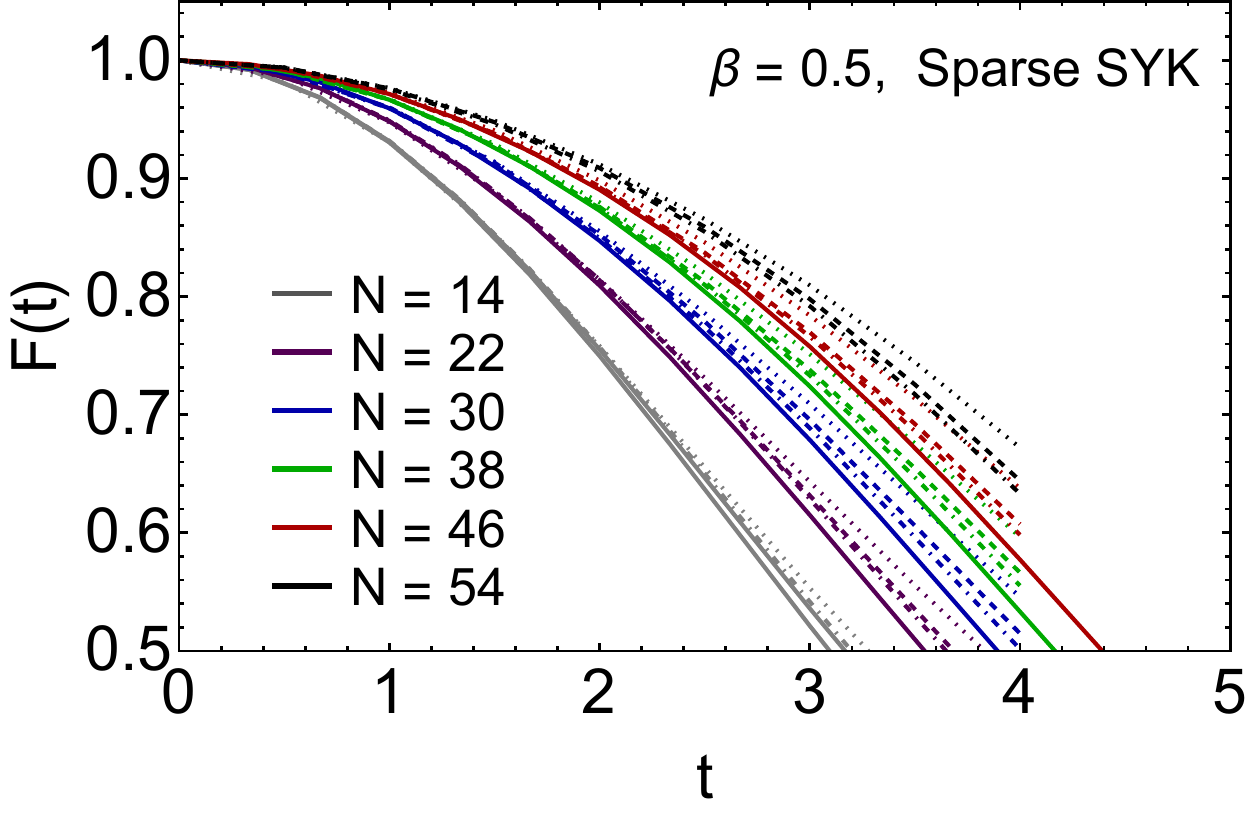}
\includegraphics[width=6cm]{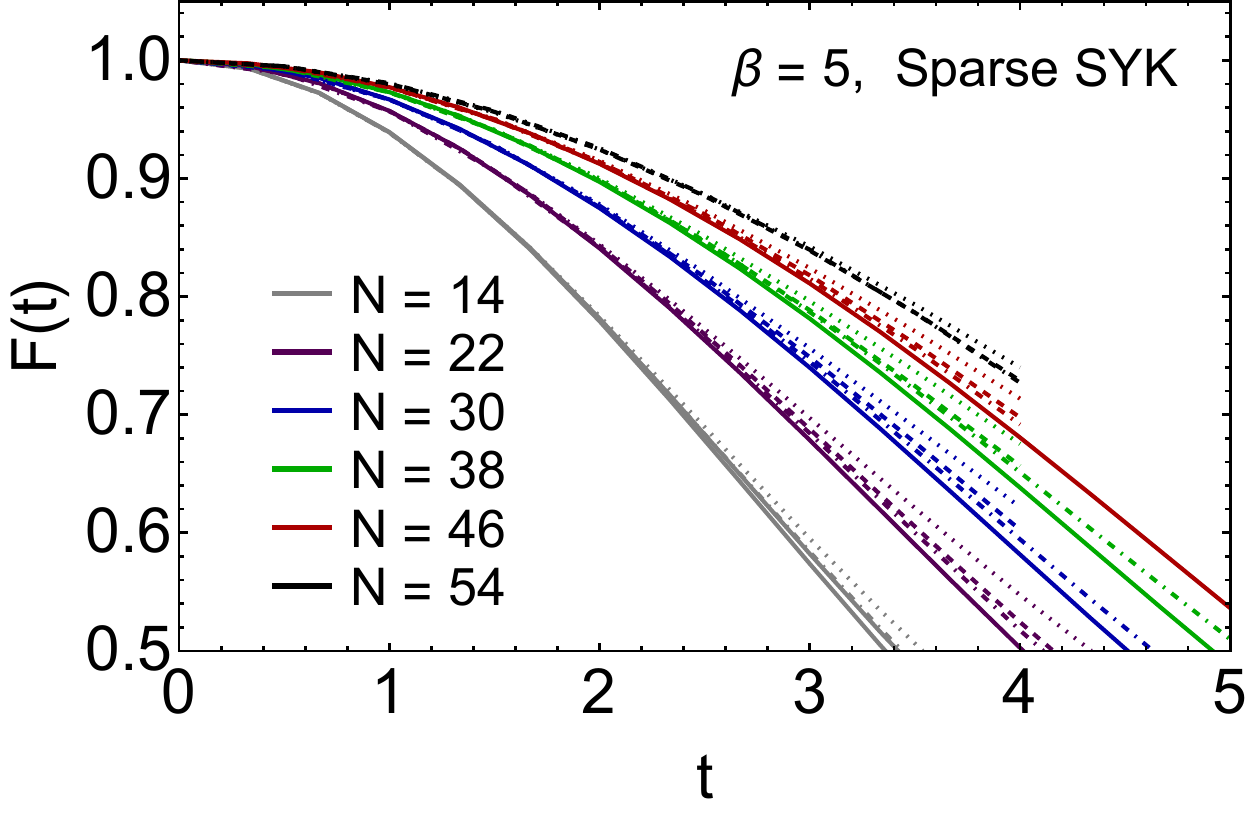}\\
\includegraphics[width=6cm]{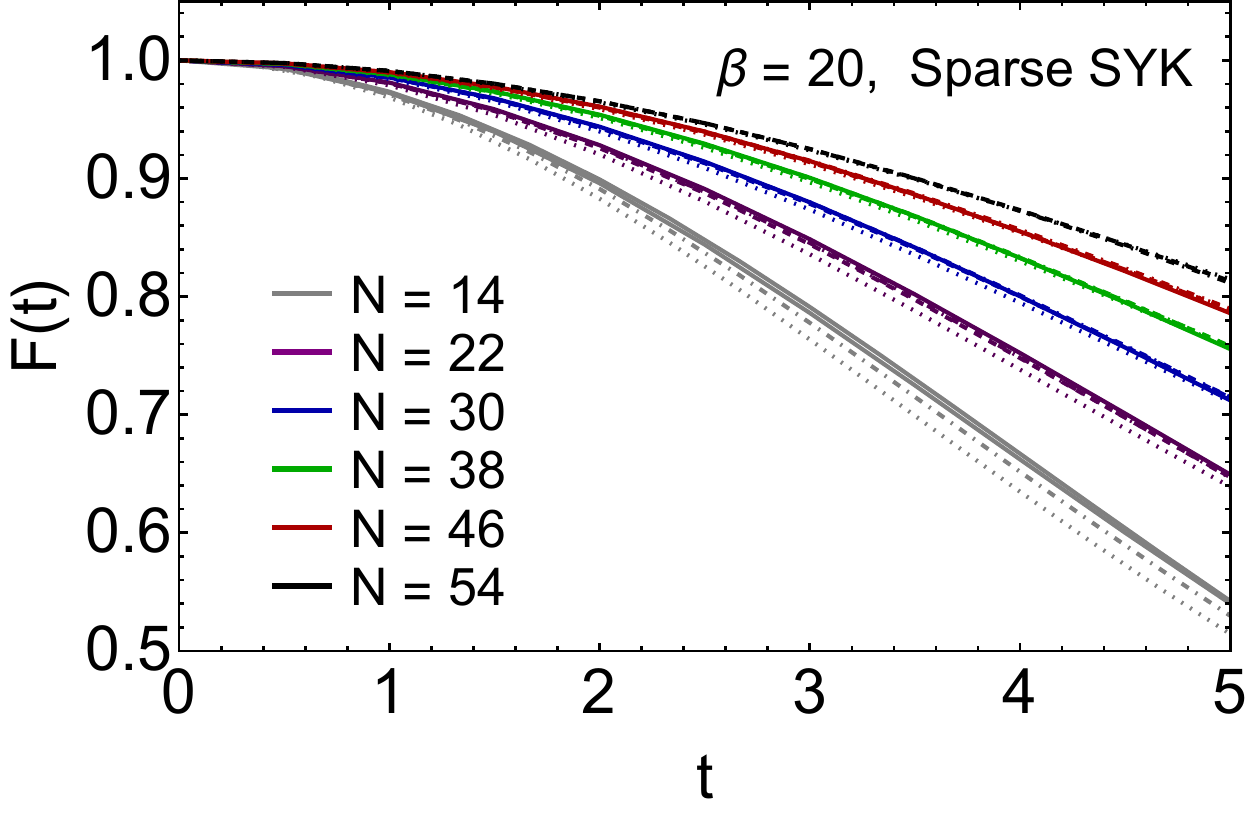}
\caption{The OTOC Eq.~(\ref{eq:otoc}) versus time $t$ for $N = 14, \; 22, \; 30,\; 38,\;46$ and 54 as indicated in the legend. Results are given for
  $\beta =0.5$ (upper), $\beta  = 5$ (middle) and $\beta = 20 $ (lower) and sparseness parameter $k=3$ (dotted), $k=6$ (dashed), $k=9$ (dot-dashed)
and dense (solid). The dependence on $k$ in the region where OTOC decreases exponentially is quite weak.}
\label{fig:otoc}
\end{figure}

A strong hint of what will be the main result of this paper, the independence of the Lyapunov exponent on the sparsity, can already be seen from Fig.~\ref{fig:otoc} depicting the OTOC dependence on the sparseness parameter $k$ for different $N \in [14,54]$: in the low-temperature limit,
the OTOC depends only weakly  on $k$ even for relatively small $k = 3$ close  
to the percolation limit where the Hamiltonian breaks up into blocks for most disorder realizations. We restrict ourselves to $N {\rm mod 8} = 6$ because the approach to the large $N$ limit may be more uniform taking advantage of the Bott periodicity of the SYK model \cite{you2016}.

\medbreak
{\noindent\it GPU-based Numerical Optimizations.}\quad The calculation of the OTOC in the large-$N$ limit of interest requires the development of a novel quantum time evolution library \cite{liuchang2023} called {\tt REAPERS}, short for ``a {\tt REA}sonably {\tt PER}formant {\tt S}imulator for qubit systems'', that implements highly-optimized matrix-free \cite{matrixfree1993} Krylov subspace methods \cite{krylov1931,krylov2007} on nVidia GPUs. Written in C++20, it provides a programming interface similar to that of {\tt dynamite} \cite{dynamite2023}, but does not depend on low-level libraries such as the Portable, Extensible Toolkit for Scientific Computation
({\tt PETSc}) \cite{balay1997,slepc2005,slepcmanual2022} for matrix operations. Instead, we start from scratch and implement optimized
 Computer Unified Device Architecture (CUDA)  kernels that compute spin operator actions on quantum states more efficiently than those provided by {\tt PETSc}, and carefully manage object allocations and de-allocations to use as little video memory as possible. The result of these performance optimizations is that we are able to simulate sparse SYK systems with $N = 62$ fermions in double precision floating point and $N = 64$ fermions in single precision, on single-GPU systems with the 80GB version of the nVidia A100 graphics card.
 Comparing this with the previous state of the art CPU-based calculation of an $N = 60$ (albeit dense) SYK system using a 500-node supercluster
 \cite{kobrin2020}, our hardware cost is far less and we consume far less energy. We refer the reader to the Supplemental Materials for the technical details of our optimization techniques and further benchmark data.

\medbreak
{\noindent\it Finite Size Scaling Analysis.}\quad The expectation for quantum chaotic systems \cite{larkin1969,berman1978} is that for sufficiently short times below the Ehrenfest time the OTOC decreases exponentially.
 At least for low temperatures, an analysis based on the Schwarzian action
\cite{maldacena2016a,bagrets2017,bagrets2016,altland2021late} in the dense case shows that the decay only remains exponential up to around the Ehrenfest time, after which it approaches
zero with a decreasing exponent before finally turning into a power-like decay for very long times. In the exponentially decaying domain, the dependence on
$t$ and $N$ is only through the combination $\exp(\lambda_L t)/N$
\cite{maldacena2016,Lam:2018pvp,Yang:2018gdb}
so that
\be\label{eq:ft}
F(t)=g_0(t)- g\left ( \frac{e^{\lambda_L t}}{N}\right )
\ee
with
\be
g\left ( \frac{e^{\lambda_L t}}{N}\right ) =c_1 \frac{e^{\lambda_L t}}{N} + c_2 \left (\frac{e^{\lambda_L t}}{N} \right)^2 + \cdots\;
\label{atoc}
\ee
and $\lambda_L$ the Lyapunov exponent.
In order to extract $\lambda_L$, we have to restrict the numerical calculation of $F(t)$ to the region in which the decay 
obeys Eq.~(\ref{atoc}).
For that purpose, we largely follow the method of Ref.~\cite{kobrin2020} for the dense SYK based on the rescaling symmetry, $t \to t + \log{r}/\lambda_L$ and $N \to Nr$ where $r > 0$.
In a first step, we determine the time $t^*(N)$ for which $F(t)$ drops to a certain value $F_0 < 1$. The value of $F_0$ cannot be too large because that would not capture the exponential growth but it also cannot be too small because the OTOC no longer decays exponentially.
We shall see that for values of $F_0$ between $0.75$ and $0.85$ the results are consistent with an exponential growth. For the scaling behavior Eq.~\eref{atoc} we find to leading order in $1/N$,
\be\label{eq:tstar}
t^*(N)= \frac {\log N}{\lambda_L} + \frac 1{\lambda_L} g^{-1}((g_0(t^*(N))-F_0)/c_1).
\ee
The rescaling symmetry requires that $g_0$ does not depend on $t$ in the region of exponential decay.  We will see in the Supplemental Material that $g_0(t)$ depends only weakly on $t$.
In that case, the second term can be eliminated by differentiating with respect to $N$, resulting
in
\be
\lambda_L= \frac 1 {N d t^*(N)/dN} + O (1/N).
\ee
In principle, the Lyapunov exponent can be obtained from the slope of $t^*(N)$
versus
$\log N$, but the slope  has a residual $N$-dependence in the time and size window at our disposal. Ideally, we fit observables for which
this residual $N$-dependence is minimized. In agreement with \cite{kobrin2020},
our numerical results suggest that the $1/N$ dependence of
the inverse slope is close to linear at low  temperatures ($\beta \ge 5$) so that the Lyapunov exponent is determined by
\be
\frac 1 {N d t^*(N)/dN}   = \lambda_L + \frac {\alpha_1} N+\frac {\alpha_2}{ N^2}
+ O (1/N^3)
\label{lamexpan}
\ee
with $\alpha_2=0$.
At high temperatures (say $\beta =0.5$), the $1/N$ dependence is fitted by a quadratic
dependence with $\alpha_1=0$ except in the dense case when the data are
sufficiently accurate to use a three
parameter fit.
An estimate for the
Lyapunov exponent is given by the extrapolation of $1/({N d t^*(N)/dN})$
to $1/N \to 0$.  
Details of the fitting procedure  are left to the Supplemental Material.

\begin{figure}[t!]
  \centering
  \includegraphics[width=5cm]{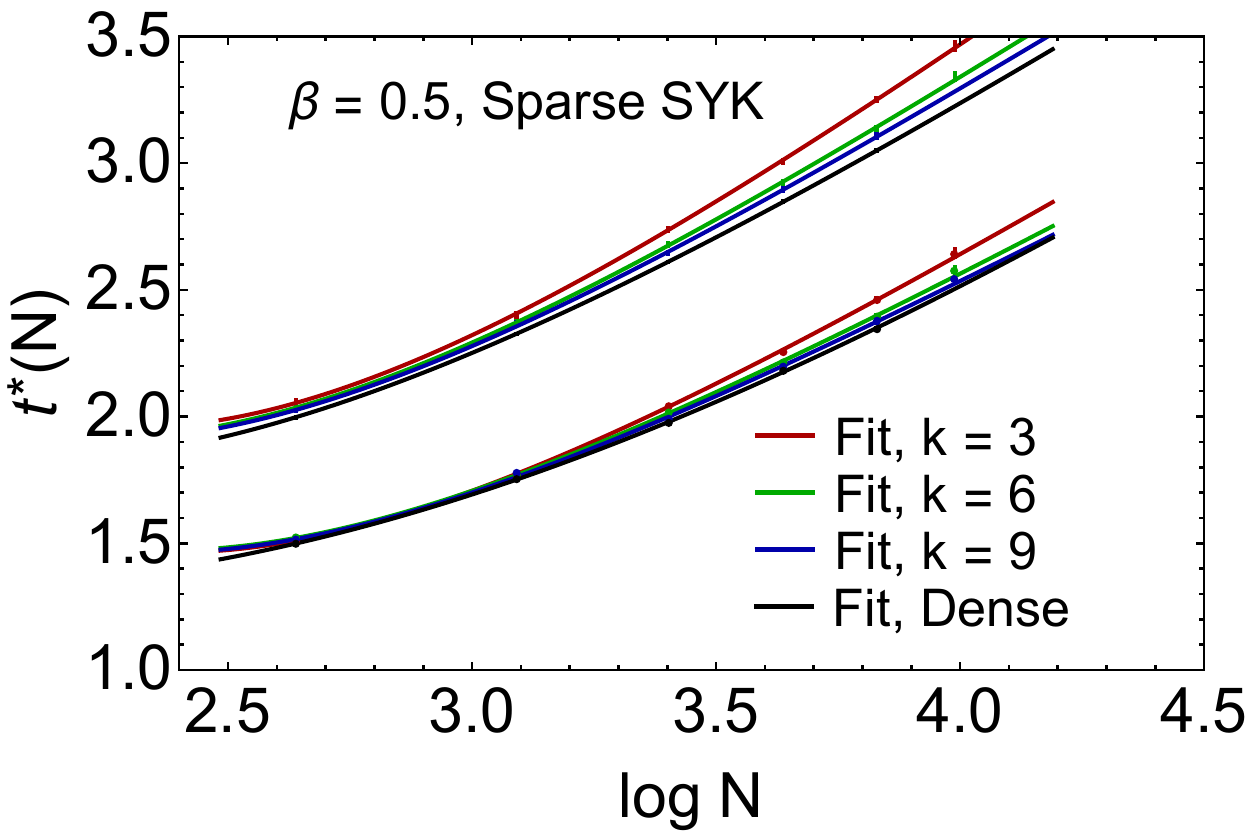}
  \includegraphics[width=5cm]{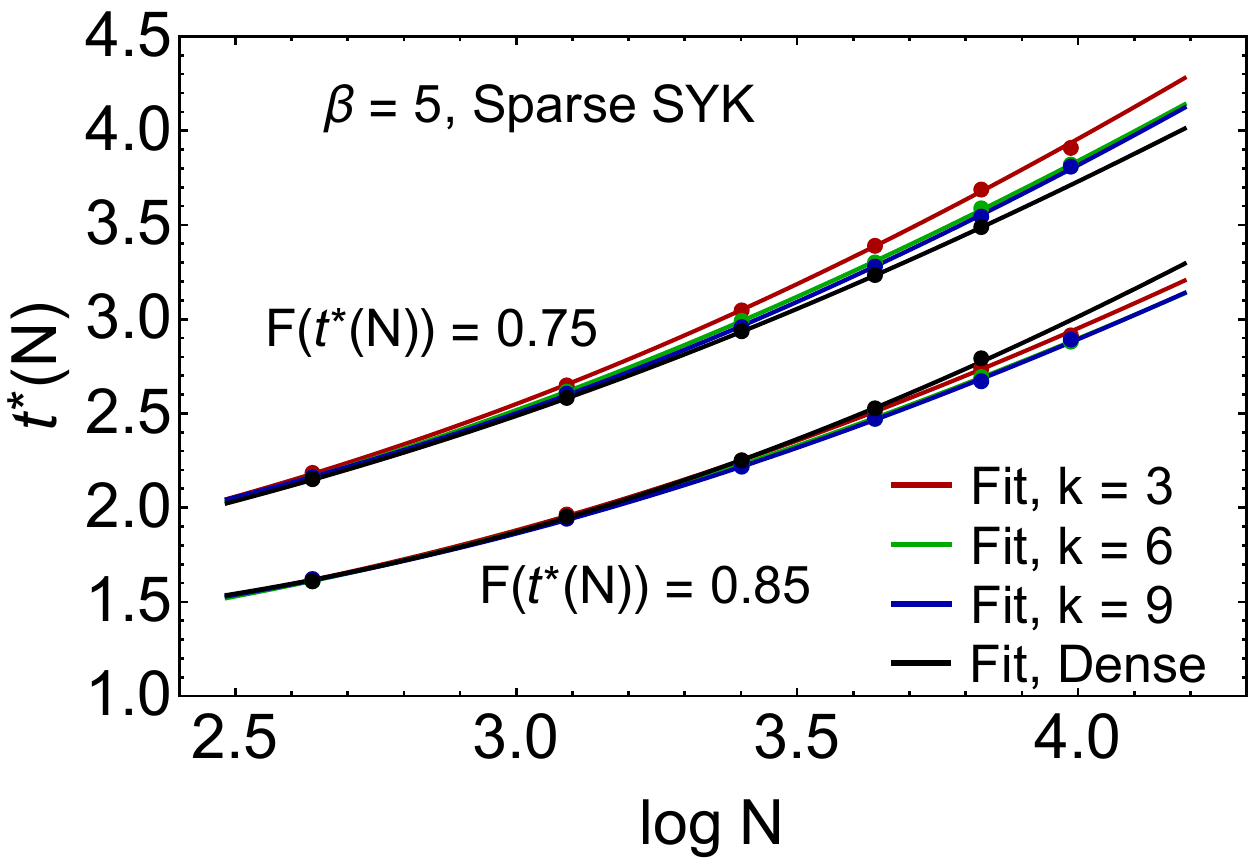}
  \includegraphics[width=5cm]{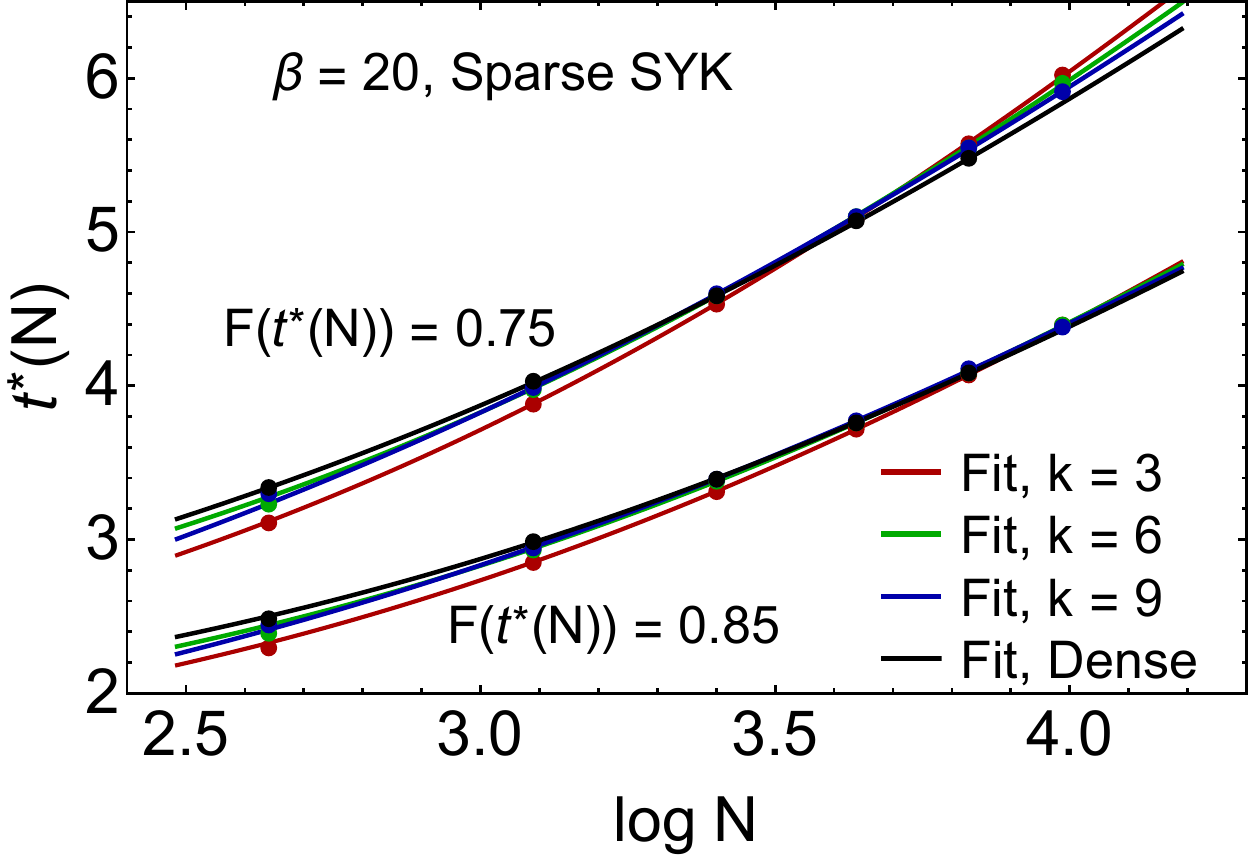}
  \caption{The dependence of $t^*(N)$ as a function of $\log N$  for $\beta =0.5$ (upper), $\beta=5$ (middle)
    and $\beta=20$ (lower) and values of $k$ as indicated in the legends of the figures. 
    For the numerical fit we employ Eq.~(\ref{glob}). For $\beta=20$, $k=3$ and $N=14$, 13 outliers that
differ by more than 30
standard deviations from the median have been excluded from the
averages \cite{rousseeuw2011robust}.}
  \label{fig:tnst}
\end{figure}
\begin{figure}[t!]
  \centering
  \includegraphics[width=6cm]{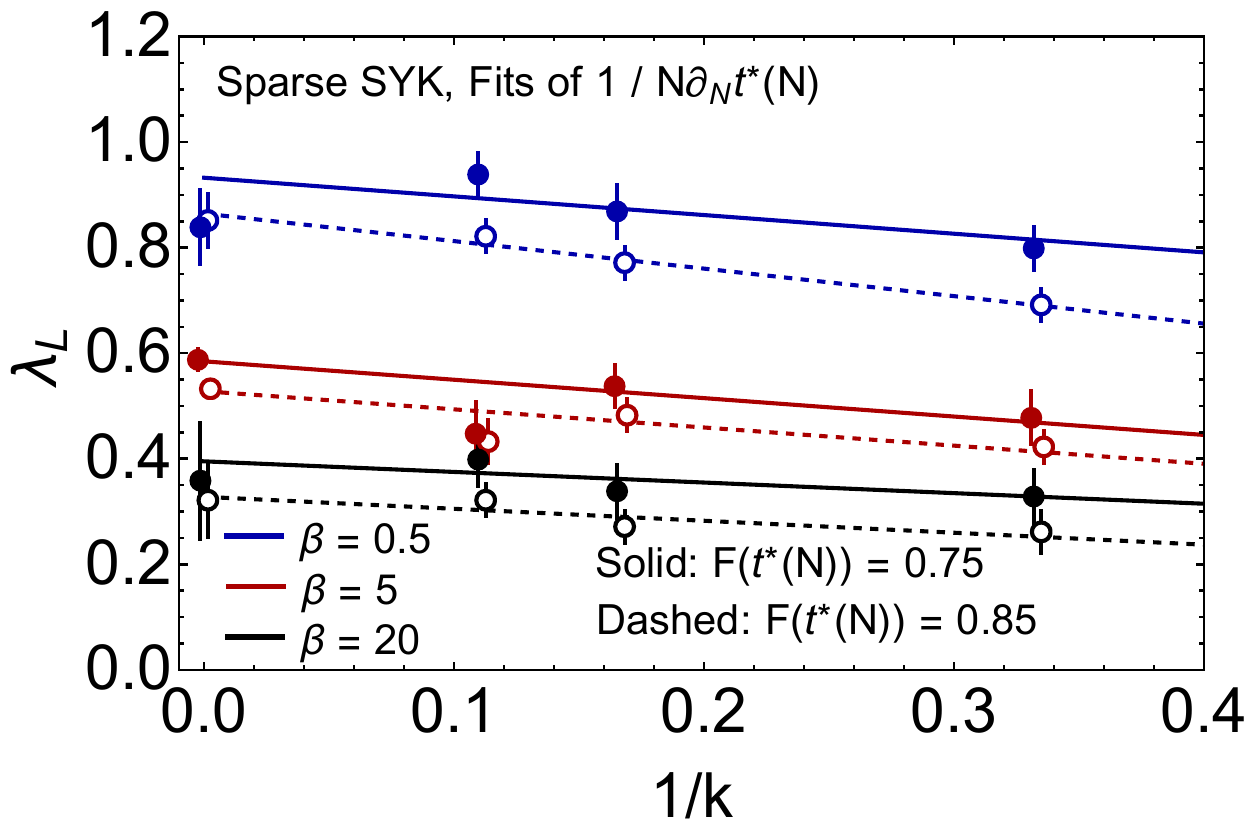}
  \includegraphics[width=6cm]{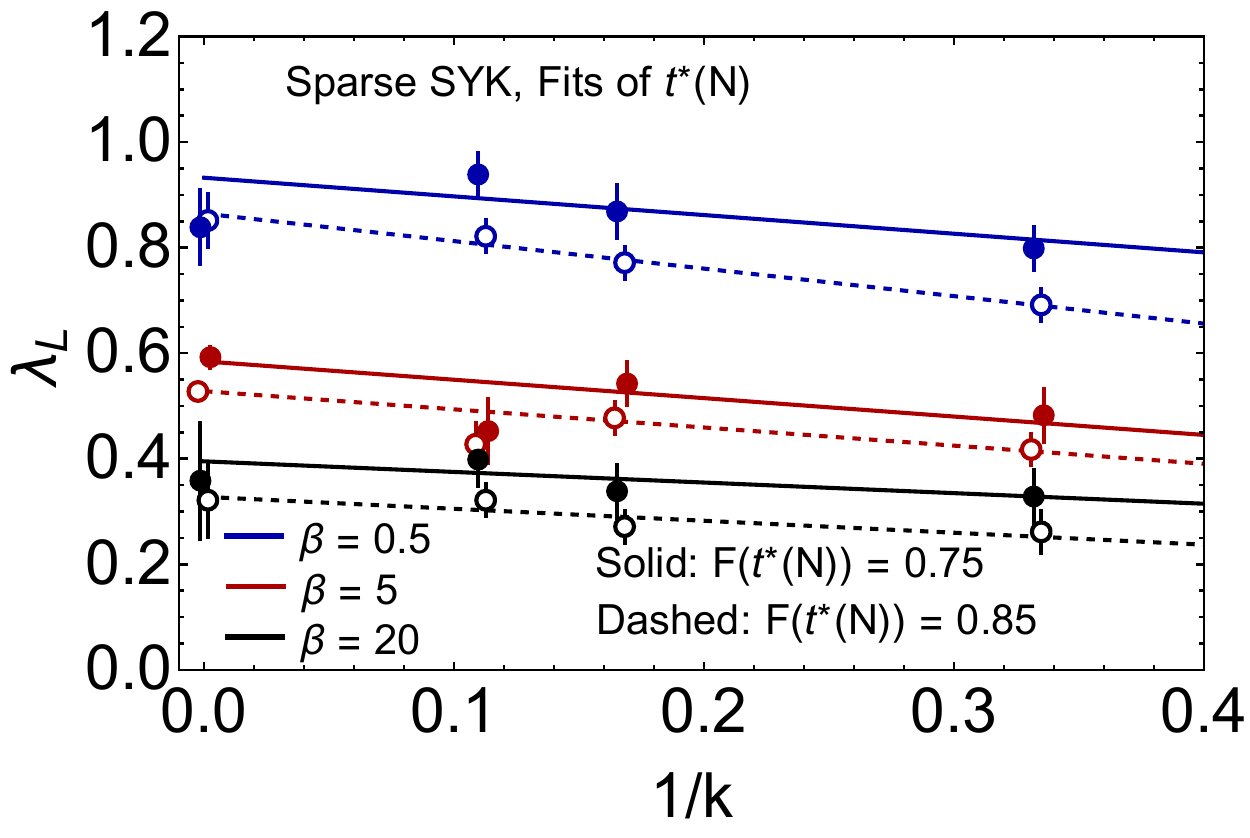}
  \caption{Top: The Lyapunov exponent, obtained from $t^*(N)$ (bottom) and
    the derivative of $t^*(N)$ (top), as a function of the inverse  sparsity parameter $k$ for $\beta = 0.5$, $5$,
    and $20$ using two different cutoffs $F_0 = 0.75$ (dashed and open circles), $0.85$ (solid and solid disks). The dense case is represented by the point $1/k = 0$. The lines stand for linear fits to the data points. See Table \ref{tab:lya} for a summary of results. 
For $\beta =20$, the slope of the line which accounts for the $k$ dependence is consistent with zero.
  }
  \label{fig:lya}
\end{figure}
\begin{table*}
  \centering
  \begin{tabular}{l|llllc}
    $\lambda_L$& $F_0=0.75,\;1/(N\del_N t^*(N)) $ &
$F_0=0.85,\;1/(N\del_N t^*(N))$ &
    $F_0=0.75,\; t^*(N) $ & $F_0=0.85,\;t^*(N)$ &Dense SD\\
    \hline
    $\beta=0.5$
    & $0.86(3)-0.52(15)/k $\hspace*{0.5cm} & $0.93(4)-0.35(20)/k $
   \hspace*{0.5cm} & $0.84(2)-0.29(10)/k$\hspace*{0.5cm} & $0.94(2)-0.20(7)/k$
    \hspace*{0.5cm}& 0.86\\
    $\beta=5$ &
    $0.52(1) -0.34(9)/k$& $0.58(2) -0.19(14)/k $
    & $0.50(1)-0.22(4)/k$ &$0.60(1)-0.22(6)/k$
    & 0.59\\
    $\beta=20$&
    $0.33(4)-0.2(2)/k$ & $0.39(6)-0.2(3)/k$
    &$0.30(2)-0.16(10)/k$& $0.40(3) -0.2(2)/k$
    &0.24\\
    \hline
  \end{tabular}
  \caption{The Lyapunov exponents obtained by a linear fit to the data
    of Fig. \ref{fig:lya}.
    In the first two columns,  $\lambda_L$ is obtained from Eq.~(\ref{glob}), in the next two columns $\lambda_L$ is obtained from Eq.~(\ref{lamexpan}). Dense SD stands for the analytic result \cite{maldacena2016}.}
  \label{tab:lya}
\end{table*}

Alternatively, one can integrate Eq.~\eref{lamexpan}, resulting in the equivalent
expansion up to logarithmic factors,
\be
t^*(N) = \gamma_0 + \frac {\log N}{\lambda_L} \left ( 1 + \frac {\gamma_1}N
+ \frac {\gamma_2}{N^2} \right )+ O \left (\frac 1{N^3}\right )
\label{glob}
\ee
where $\lambda_L,\; \gamma_0$,  $\gamma_1$ and $\gamma_2$ are fitting parameters. For $\beta\ge 5$ we set $\gamma_2 =0$, but
for $\beta = 0.5$ we use $\gamma_2/N^2$ as the correction term putting
$\gamma_1=0$. As can be seen from Fig. \ref{fig:tnst}, this gives an excellent
fit of the $t^*(N)$ data
for all considered temperatures, sparsity parameters and cut-off values.
Fitting  $t^*(N)$ directly has the advantage that the errors are smaller.
On the other hand, fitting the numerical derivative $\del_N t^*(N)$ has
the benefit of having one less fitting parameter at the expense of
much larger errors (see Supplemental Materials). We shall see the fitting results of both methods are consistent
though there is a
significant systematic error.
There is also an issue
of over-fitting which trades the $\log N$ dependence for the $1/N$ dependence.
For example using additive $1/N$ corrections
instead of multiplicative $1/N$ correction
significantly changes the  value of the Lyapunov exponent.

\medbreak
{\noindent\it Results for the Lyapunov Exponent.}\quad Our results for the Lyapunov exponents $\lambda_L$ are shown in Fig.~\ref{fig:lya}.  In the range of temperatures we have considered,  $\beta = 0.5,$ $5$, $20$, there is no significant dependence of the Lyapunov exponent on the sparsity parameter $k$. The difference of $\lambda_L$ for different values of $k$ is less
than the discrepancies between
the two choices of $F_0$ and the two choices of the fitting methods,
both of which are a measure of
the systematic error.
Numerical results for the fitted lines are given in Table \ref{tab:lya}.

We now compare our results with predictions for the dense case, either analytical or based on the numerical solutions of the Schwinger-Dyson (SD) equations. As shown in the last column of Table \ref{tab:lya}, for $\beta =0.5$ and $\beta=5$ our numerical results
are in good agreement with the large-$N$ prediction obtained by solving the
SD equations \cite{kobrin2020}. For $\beta=20$, we find a Lyapunov exponent that is consistent
with the chaos bound of $2\pi/\beta$ \cite{maldacena2015}. Taking
into account subleading finite temperature corrections (which can be obtained
by solving the SD equations)
lowers the theoretical large-$N$ value to 0.24, which is still above our result but
is in agreement with previous numerical calculations in the dense case \cite{kobrin2020}, where for $\beta =17.8$
the Lyapunov exponent was 0.36 versus 0.26 from solving the SD equations.
Therefore this discrepancy is not related to the sparsity of the model.
We stress that in order to reach this relatively low level of statistical fluctuations, it is necessary to simulate a number of disorder realizations at least of order $10^4$, which is several orders of magnitude larger than in the dense case \cite{caceres2023}. Surprisingly, unlike the dense case, the fluctuations are not larger for low temperature and they are not reduced as $N$ increases, which prevents us from including $N = 62$ in this analysis despite the fact that it is numerically accessible.
\medbreak
{\noindent\it Conclusions and outlook.}\quad We have studied out-of-time-order correlators in a sparse variant of the SYK model. After some careful data analysis, we have shown that the Lyapunov exponent has no significant dependence on the sparsity $k$ for all temperatures we have considered, and agrees with previous \cite{kobrin2020} numerical results for the dense case. In the low temperature limit, the value of the Lyapunov exponent for both the dense and  the sparse SYK is above the analytical prediction \cite{maldacena2016}, which prevents us from explicitly confirming that the bound on chaos \cite{maldacena2015} is still saturated for the sparse SYK model. However, the independence of the Lyapunov exponent on the sparsity indicates that the low-temperature bound will also be saturated in the sparse case.

A crucial part of our work is the development of an optimized GPU-based quantum simulation library, which enables us to reach $N \leq 64$ Majoranas due to drastic improvements in simulation speed and memory usage. The energy consumption and the cost of hardware are both vastly smaller than equivalent simulations on CPU-based systems.
Natural extensions of this work includes
computing OTOCs in non-Hermitian SYK and sparse spin chains, such as those employed in studies of many-body localization \cite{basko2006,luitz2015}.
\medbreak
\begin{acknowledgments}
AMGG and CL were partially supported by a National Key R$\&$D Program of China (Project ID: 2019YFA0308603), and a Shanghai talent program. JJMV acknowledges support from U.S.\ DOE Grant No. DE-FAG88FR40388.
\end{acknowledgments}

\bibliography{library}

\ifarXiv
    

\section*{Supplementary material}

# Supplemental Materials: Saturation of the bound on chaos in sparse Sachdev-Ye-Kitaev models

In the first section of the supplemental material we will discuss details of the numerical algorithms and performances tests for the calculation of OTOCs for the sparse and dense SYK model. Details of the finite size scaling analysis are given in the second section.

## I. ALGORITHMS AND PERFORMANCE TESTS

The numerical simulations in this work use the standard methodology where we represent the SYK Majorana fermions as tensor products of spin-1/2 operators. These Majorana fermion field operators are block off-diagonal and satisfy $\{\gamma_i, \gamma_j\} = 2\delta_{ij}$. The explicit expression is

$$\gamma_n = \begin{pmatrix} 0 & \gamma_n^{\text{UD}} \\ \gamma_n^{\text{DU}} & 0 \end{pmatrix}, \qquad n = 0, 1, \ldots, N-1, \qquad N \quad \text{even},$$

where

$$\gamma_0^{\text{UD}} = \gamma_0^{\text{DU}} = \sigma_y \bigotimes_{0}^{N/2-2} \sigma_x,$$

$$\gamma_{2n+1}^{\text{UD}} = \gamma_{2n+1}^{\text{DU}} = \sigma_y \left( \bigotimes_{\lfloor (n+1)/2 \rfloor}^{N/2-2} \sigma_x \right) \left( \bigotimes_{0}^{\lfloor (n-1)/2 \rfloor} \mathbb{1} \right), \quad n = 1, \ldots, N-2,$$

$$\gamma_{2n}^{\text{UD}} = \gamma_{2n}^{\text{DU}} = \sigma_z \left( \bigotimes_{\lfloor (n+1)/2 \rfloor}^{N/2-2} \sigma_x \right) \left( \bigotimes_{0}^{\lfloor (n-1)/2 \rfloor} \mathbb{1} \right), \quad n = 1, \ldots, N-2, \qquad (S1)$$

and

$$\gamma_{N-1}^{\text{UD}} = -\gamma_{N-1}^{\text{DU}} = -i\mathbb{1}.$$

The Pauli matrices are defined as

$$\sigma_0 = \mathbb{1} = \begin{pmatrix} 1 & 0 \\ 0 & 1 \end{pmatrix}, \qquad \sigma_1 = \sigma_x = \begin{pmatrix} 0 & 1 \\ 1 & 0 \end{pmatrix},$$

$$\sigma_2 = \sigma_y = \begin{pmatrix} 0 & i \\ -i & 0 \end{pmatrix}, \qquad \sigma_3 = \sigma_z = \begin{pmatrix} 1 & 0 \\ 0 & -1 \end{pmatrix}.$$

This notation can be extended to spin systems. For example, the spin operator at site $i$ is given by

$$\sigma_x^{(i)} = \mathbb{1} \otimes \cdots \otimes \underbrace{\sigma_x}_{i\text{-th site}} \otimes \cdots \otimes \mathbb{1}$$

We sometimes write $\sigma_x(i)$ in lieu of $\sigma_x^{(i)}$ when the superscript is hard to read. Using this notation eq. (S1) can be written in the following form

$$\gamma_n^{\text{UD}} = \gamma_i^{\text{DU}} = \sigma_{y/z}\left(\left\lfloor \frac{n-1}{2} \right\rfloor\right) \prod_{i=\lfloor \frac{n+1}{2} \rfloor}^{N/2-2} \sigma_x(i), \quad n = 0, 1, \ldots, N-2$$

where the $\sigma_{y/z}\left(\left\lfloor \frac{n-1}{2} \right\rfloor\right)$ factor stands for the following expression

$$\sigma_{y/z}\left(\left\lfloor \frac{n-1}{2} \right\rfloor\right) = \begin{cases} \mathbb{1} & n = 0 \\ \sigma_y\left(\left\lfloor \frac{n-1}{2} \right\rfloor\right) & n \text{ is odd} \\ \sigma_z\left(\left\lfloor \frac{n-1}{2} \right\rfloor\right) & n \text{ is even} \end{cases}$$

The OTOCs and Green functions are then computed by first constructing a Haar-random state, then evolving the state using the same Krylov subspace method described in [S1], and finally computing the inner product of the initial state with the final state. Disorder averaging is then performed over all disorder realizations. The explicit expression we have used to compute the Green function is

$$G_{\rm R}(t) = {\rm Re}\,\frac{G(t)}{G(0)}$$

where

$$G(t) = \langle\psi|e^{(it-\beta/2)H}\gamma_{N-1}e^{-iHt}\gamma_{N-1}e^{-\beta H/2}|\psi\rangle$$

which in terms of parity blocks is

$$G(t) = \langle\psi_{\rm U}|e^{(it-\beta/2)H_{\rm UU}}\gamma_{N-1}^{\rm UD}e^{-iH_{\rm DD}t}\gamma_{N-1}^{\rm DU}e^{-\beta H_{\rm UU}/2}|\psi_{\rm U}\rangle + \{\text{swap UD}\}$$

Here, $|\psi\rangle$ is a Haar-random state and $|\psi_{\rm U}\rangle$ and $|\psi_{\rm D}\rangle$ are its parity components. For the OTOC we have

$$\text{OTOC} = \frac{F(t)}{F(0)}$$

where

$$F(t) = \langle\psi|e^{(it-\beta/8)H}\gamma_{N-1}e^{-(it+\beta/4)H}\gamma_{N-2}e^{(it-\beta/4)H}\gamma_{N-1}e^{-(it+\beta/4)H}\gamma_{N-2}e^{-\beta H/8}|\psi\rangle$$

which in terms of parity blocks is given by

$$\begin{aligned}F(t) =& \langle\psi_{\rm U}|e^{(it-\beta/8)H_{\rm UU}}\gamma_{N-1}^{\rm UD}e^{-(it+\beta/4)H_{\rm DD}}\gamma_{N-2}^{\rm DU}e^{(it-\beta/4)H_{\rm UU}}\gamma_{N-1}^{\rm UD}e^{-(it+\beta/4)H_{\rm DD}}\gamma_{N-2}^{\rm DU}e^{-\beta H_{\rm UU}/8}|\psi_{\rm U}\rangle \\ &+ \text{swap UD}\end{aligned}$$

Note that we have explicitly chosen the last and second to last $\gamma$ matrices in the expressions for the Green function and the OTOC. This is due to the fact that

$$\gamma_{N-1}^{\rm UD} = -\gamma_{N-1}^{\rm DU} = -i\mathbb{1}$$

which allows us to eliminate the matrix multiplications involving $\gamma_{N-1}$. For $\gamma_{N-2}$ we have

$$\gamma_{N-2}^{\rm UD} = \gamma_{N-2}^{\rm DU} = \sigma_z^{(N/2-2)}$$

Using the two lines of expressions above the OTOC simplifies to

$$\begin{aligned}F(t) =& \langle\psi_{\rm U}|e^{(it-\beta/8)H_{\rm UU}}e^{-(it+\beta/4)H_{\rm DD}}\sigma_z^{(N/2-2)}e^{(it-\beta/4)H_{\rm UU}}e^{-(it+\beta/4)H_{\rm DD}}\sigma_z^{(N/2-2)}e^{-\beta H_{\rm UU}/8}|\psi_{\rm U}\rangle \\ &+ \text{swap UD}\end{aligned}$$

The main numerical effort is focused on optimizing this calculation on nVidia Graphical Processing Units (GPUs). Previous work [S1, S2] relied on a Python package called `dynamite` [S3], which is ultimately a simple front end to the underlying C libraries that implement the basic matrix operations (`PETSc`, Portable, Extensible Toolkit for Scientific Computation) and on top of that, the Krylov algorithms (`SLEPc`, Scalable and Flexible Toolkit for the Solution of Eigenvalue Problem). In the early stage of our work, we successfully reproduced the claim in Ref. [S1] using `dynamite` that the dense SYK model with $N = 50$ fermions can be simulated (within a reasonably amount of time) on a single nVidia V100 graphics card. However, when examining the performance of these simulations, we realized that the software is sub-optimal in terms of both speed and memory: the observed power usage is consistently below half of the nominal maximum power of the graphics cards, and the video Random Access Memory (RAM) usage is far above the theoretical minimum. To elaborate on the latter, for the matrix-free representation of the spin operators (called 'shell matrix' by `dynamite`), the theoretical minimum RAM usage is $(v + 3) * ($ `size of state vector` $)$, where $v$ is the dimension of the Krylov subspace and the size of the state vector is $(\text{fp}/4) \cdot 2^{N/2-1}$ with fp being the floating point precision (fp = 32 or 64). The $v+3$ vectors consist of $v+1$ vectors used during the Krylov algorithm (including the initial state vector), plus a vector to store the final result of the Krylov evolution, plus a copy of the initial state vector for the purpose of computing the inner product (this last vector can be 'swapped' to host memory temporarily and then reloaded from there in order to save video memory usage, without a significant loss of performance). In our experiment, `dynamite` and its software stack consumed far more video memory than this theoretical minimum.

The issue with the performance is ultimately due to the underlying `PETSc` library which has a huge, ancient code base, one that was first designed for Central Processing Unit (CPU) computing and later ported to different parallel computing architectures, including General Purpose Graphical Processing Units (GPGPU). For those unfamiliar with the code base, any further performance work would most likely cost more engineering time than simply starting from scratch. Seeing that there is presently a lack of high-performance, GPU-enabled quantum spin model simulators, we have therefore decided to write a C++ library that presents a similar `dynamite`-like interface, but with GPU computing as a first class citizen from day one. We focus on nVidia Computer Unified Device Architecture (CUDA) as it provides the most mature GPGPU ecosystem, but the library is designed to be portable to different parallel computing architectures. In fact, when CUDA is not available, the library falls back to CPU-based parallel computing using OpenMP. The library provides basic primitives such as spin operators and quantum state vectors, and on top of these we re-implement the same restarted Krylov algorithm [S4] as in `SLEPc` for computing the state time evolution.

The next part of the Supplementary Material will focus on the specifics of the performance work that we have alluded to in the previous paragraph. Once that is explained, we present the performance benchmark in the context of SYK simulations using our library code. We refrain from publishing speed comparisons with `dynamite` as this is inherently biased, and will simply present the measured computational times in our numerical simulations. Due to differences of hardware configurations, compiler directives, compiler versions, simulation parameters, etc., results may look different from these numbers.

### A. Optimization Techniques

The key to a high performance implementation of quantum spin model simulators is finding an efficient encoding of spin operators in such a way that its operations map to bit-level hardware primitives effectively. `dynamite` uses what it calls the Mask-Sign-Coefficient (MSC) representation of a spin operator which uses a list of 3-tuples, namely 'mask', 'sign', and 'coefficient' to represent a spin operator expanded in the Pauli basis. Our representation is inspired by this, but we only use the 'mask' and 'coefficient' and eliminate the 'sign' part of the tuple as it proves to be unnecessary. Our representation is as follows: for a spin operator defined on a spin-1/2 chain of length $L$, the tensor products of Pauli matrices form a basis for the space of operators. In other words, for any spin operator $S$ we can always expand it as

$$S = \sum_{i_0, i_1, \ldots, i_{L-2}, i_{L-1}} \lambda_{i_{L-1} i_{L-1} \ldots i_1 i_0} \, \sigma_{i_{L-1}} \otimes \sigma_{i_{L-2}} \otimes \cdots \otimes \sigma_{i_1} \otimes \sigma_{i_0}$$

where each $i_k = 0, 1, 2, 3$. For a single term in the sum, we use a 2-tuple, consisting of a bit-string which we will call `bits` in what follows, and a complex number which we will call `coeff`. The term $\lambda_{i_{L-1} i_{L-2} \ldots i_1 i_0} \, \sigma_{i_{L-1}} \otimes \sigma_{i_{L-2}} \otimes \cdots \otimes \sigma_{i_1} \otimes \sigma_{i_0}$ will be represented as $\langle \text{bits}, \text{coeff} \rangle$ where

$$\text{bits} = s_{i_{L-1}} s_{i_{L-2}} \cdots s_{i_1} s_{i_0}$$
$$\text{coeff} = \lambda_{i_{L-1} i_{L-2} \ldots i_1 i_0} \, i^{\#\sigma_y}$$

Here $s_{i_{L-1}} s_{i_{L-2}} \cdots s_{i_1} s_{i_0}$ is a concatenation of bit strings $s_i$ where $s_0 = 00$, $s_1 = 01$, $s_2 = 11$, and $s_3 = 10$, and $\#\sigma_y$ is the number of $\sigma_y$'s in $\{\sigma_{i_0}, \ldots, \sigma_{i_{L-1}}\}$. Note that all the bit-strings in this paper have the most significant bit in the left-most place.

The reason for encoding the $\sigma$ matrices in this fashion should be apparent once the reader realizes that the action of $\sigma_x$ on a spin site simply flips the qubit (ie. exchange the amplitudes for $|0\rangle$ and $|1\rangle$), and the action of $\sigma_z$ simply negates the amplitude for $|1\rangle$. Ignoring the factor of $i$, the action of $\sigma_y$ is simply that of $\sigma_x$ followed by $\sigma_z$. The factors of $i$ can be absorbed into the overall coefficient $\lambda_{i_{L-1} i_{L-2} \ldots i_1 i_0}$, because tensor products commute with scalar multiplications. Therefore to compute the action of a single term $\lambda_{i_{L-1} i_{L-2} \ldots i_1 i_0} \, \sigma_{i_{L-1}} \otimes \sigma_{i_{L-2}} \otimes \cdots \otimes \sigma_{i_1} \otimes \sigma_{i_0}$ on a state vector $|\Psi\rangle = \sum_{n \in \{0,1\}^L} \Psi_n |n\rangle$, we simply scan the bit-string `bits`, starting from the least-significant bit, and whenever we see a 1 in an even-numbered position, we flip the corresponding qubit, and whenever we see a 1 in an odd-numbered position, we negate the corresponding amplitude $\Psi_n$, and finally we multiply the resulting state vector by the overall coefficient `coeff`.

This leads to Algorithm 1 for computing the action of a generic spin operator $S$ on a state $|\Psi\rangle$. Note that for this algorithm, due to the commutativity of additions (subject to the loss of precision of floating point arithmetic) and a lack of explicit data dependency, all three for-loops, including the inner-most loop, can be evaluated without regard to the order in which their bodies are executed. This algorithm is therefore "trivially parallel", in the sense that one can simply evaluate each iteration of a given loop in multiple computing units simultaneously, and then aggregate the results. Additionally, the compiler is free to employ loop unrolling (especially for the inner-most loop), loop interchange (for the two outer loops), vectorization (including vectorization of bit operations), etc., for better

---

**Algorithm 1:** Action of spin operator $S$ on state $|\Psi\rangle$

---

**input** : Dimension of the spin chain length $L > 0$
Spin operator $S$, expressed as a list $\hat{S}$ of $\langle$`bits`, `coeff`$\rangle$ tuples.
Initial state vector $|\Psi\rangle$, expressed as a $2^L$ dimensional array $\Psi[n]$.
**output:** Final state vector $|\Phi\rangle = S|\Psi\rangle$.
**for** $0 \leq m < 2^L$ **do**
    **for each** $\langle$`bits`, `coeff`$\rangle$ *tuple in list* $\hat{S}$ **do**
        $n \leftarrow m$;
        `minus` $\leftarrow$ **false**;
        **for each** *i-th non-zero bit of* `bits` **do**
            **if** *i is even* **then**
                | flip the $i$-th bit of $n$
            **else if** *the i-th bit of n is set* **then**
                | flip `minus`
            **end**
        **end**
        **if** `minus` *is* **false** **then**
            | $\Phi[m]$ += `coeff` * $\Psi[n]$
        **else**
            | $\Phi[m]$ −= `coeff` * $\Psi[n]$
        **end**
    **end**
**end**

---

cache locality and to maximize the utility of hardware units. Furthermore, this algorithm is especially suited for sparse Hamiltonians, defined as Hamiltonians with fewer than $2^L$ terms when expanded in the Pauli basis, due to the fact that the middle loop ("for each tuple in list") loops over a relatively small list $\hat{S}$. The sparse SYK model we considered in the main text satisfies this property since the terms in the Hamiltonian scales with $N$ linearly, and therefore can be computed using this algorithm in a highly optimized fashion. In fact, the advantage of Algorithm 1 becomes greater as the terms in the Hamiltonian become fewer, so we would expect even greater performance gains when considering nearest-neighbor models such as the Ising model.

The advantage of the bit-level encoding of $\sigma$ matrices is made even more evident when we compute the product of two spin operators. Note that up to an overall minus sign, the product of two $\sigma$ matrices happens to coincide with the XOR (exclusive-OR) of their corresponding bit strings. Therefore, to compute the product of two spin operators expressed as a list of $\langle$`bits`, `coeff`$\rangle$ tuples, we first simply XOR the `bits` from either list and multiply the corresponding `coeff`'s. The overall minus sign is taken care of by scanning the `bits` and pick out those cases that produce a minus sign (this can be done either via simple pattern matching or through a lookup table). The results are then aggregated to construct the final answer. Finally, for spin operator addition, we simply merge the two lists, taking into account possible duplications of the `bits` field.

Building on Algorithm 1, we have reimplemented the restarted Krylov subspace method [S4] for computing the state time evolution $e^{-iHt}|\Psi\rangle$. This algorithm is the exact same one implemented in `SLEPc` and invoked by `dynamite`, but we take care to manage (video) memory allocations and deallocations in order to minimize both the amount of video memory used as well as the number of memory allocation and deallocation calls. This is enabled by the move semantics of the C++ language, which allows one to eliminate unnecessary copies of objects. We are therefore able to achieve a much larger $N$ than the previous state of the art, due to the significantly higher performance and lower memory usage than `dynamite`. We will present the performance benchmarks in the next section.

### B. Performance Benchmarks

Our computing environment consists of single-GPU systems equipped with either a 80GB nVidia A100 graphics card, or a 24GB nVidia RTX4090 graphics card. The A100 systems have the following hardware and software configuration:

| Component | Specification |
|---|---|
| CPU | Intel(R) Xeon(R) Platinum 8362, 32C64T @ 2.80GHz |
| GPU | nVidia A100 80GB PCIE |
| Host RAM | 1TB |
| OS | Ubuntu 22.04.1 |
| Host Compiler | Intel oneAPI DPC++/C++ Compiler 2023.1.0 |
| Device Compiler | CUDA 12.1, nVidia Driver 530.30.02 |

The RTX4090 systems have the following hardware and software configurations:

| Component | Specification |
|---|---|
| CPU | AMD EPYC 7282, 16C32T @ 3.20GHz |
| GPU | nVidia RTX4090 24GB PCIE |
| Host RAM | 256GB |
| OS | CentOS 7.9 |
| Host Compiler | GCC 11.2 |
| Device Compiler | CUDA 12.1, nVidia Driver 530.30.02 |

The following performance data are collected on the RTX4090 systems. In general, due to a combination of factors, including the card being a newer generation, we find that the RTX4090 systems perform approximately 20% faster than the A100 systems despite being a consumer-grade card. For the sparse case, the calculation time has an almost linear dependence on the sparsity $k$, so to estimate the computing time for $k = 6$, simply multiply the relevant $k = 3$ times by a factor of two. All disorder realizations have $t_{\max} = 10$ with 10 time segments (ie. 11 data points to compute, since $t = 0$ is needed for the normalization of the OTOC). All simulations are done using double precision floating point numbers and the dimension of the Krylov space is set to five.

| $N$ | $k$ | $\beta = 0.05$, Avg.Time/curve | $\beta = 0.05$, Avg.Time/curve/point | $\beta = 20$, Avg.Time/curve | $\beta = 20$, Avg.Time/curve/point |
|---|---|---|---|---|---|
| 14 | 3 | 173ms | 15.7ms | 192ms | 17.5ms |
| 14 | dense | 399ms | 36.3ms | 204ms | 18.5ms |
| 18 | 3 | 198ms | 18.0ms | 233ms | 21.2ms |
| 18 | dense | 335ms | 30.5ms | 360ms | 32.7ms |
| 22 | 3 | 358ms | 32.5ms | 421ms | 38.3ms |
| 22 | dense | 848ms | 77.0ms | 872ms | 79.2ms |
| 26 | 3 | 423ms | 38.5ms | 497ms | 45.2ms |
| 26 | dense | 2827ms | 257ms | 2782ms | 253ms |
| 30 | 3 | 488ms | 44.4ms | 578ms | 52.5ms |
| 30 | dense | 14.7s | 1337ms | 16.0s | 1454ms |
| 34 | 3 | 651ms | 59.2ms | 786ms | 71.5ms |
| 34 | dense | 147.8s | 13.44s | 171.7s | 15.61s |
| 38 | 3 | 1122ms | 102.0ms | 1427ms | 129.7ms |
| 38 | dense | 491.1s | 44.65s | 579.9s | 52.72s |
| 42 | 3 | 3946ms | 358.7ms | 4933ms | 448.5ms |
| 42 | dense | 52m 26s | 286s | 63m 3s | 344s |
| 46 | 3 | 19s | 1.77s | 24s | 2.18s |
| 46 | dense | 5h 30m | 29m 55s | 6h 36m | 36m 5s |
| 50 | 3 | 83s | 7.51s | 112s | 10.16s |
| 50 | dense | 33h | 3h | 40h | 3h 40m |
| 54 | 3 | 421s | 38.30s | 540s | 49.11s |
| 54 | dense | 193h | 17h | Estimated $> 240$ h | Estimated $> 22$ h |

The following large $N$ performance data are collected on the A100 systems. Due to the memory limitation of the RTX4090 there are no comparable data for the RTX4090 systems. These simulations have different $t_{\max}$, numbers of time segments, and floating point precisions.

| $N$ | $\beta$ | $k$ | $t_{\max}$ | #TimeSegs | Krylov dim | FP | Avg.Time/curve | Avg.Time/curve/point |
|---|---|---|---|---|---|---|---|---|
| 62 | 0.05 | 3 | 4 | 8 | 2 | 64 | 2h 45m | 18m 3s |
| 62 | 0.05 | 3 | 8 | 16 | 2 | 64 | 9h 26m | 33m 16s |
| 64 | 0.05 | 3 | 5 | 10 | 2 | 32 | 312m 43s | 28m 26s |
| 64 | 20 | 3 | 16 | 8 | 2 | 32 | 11h 45m | 1h 18m |

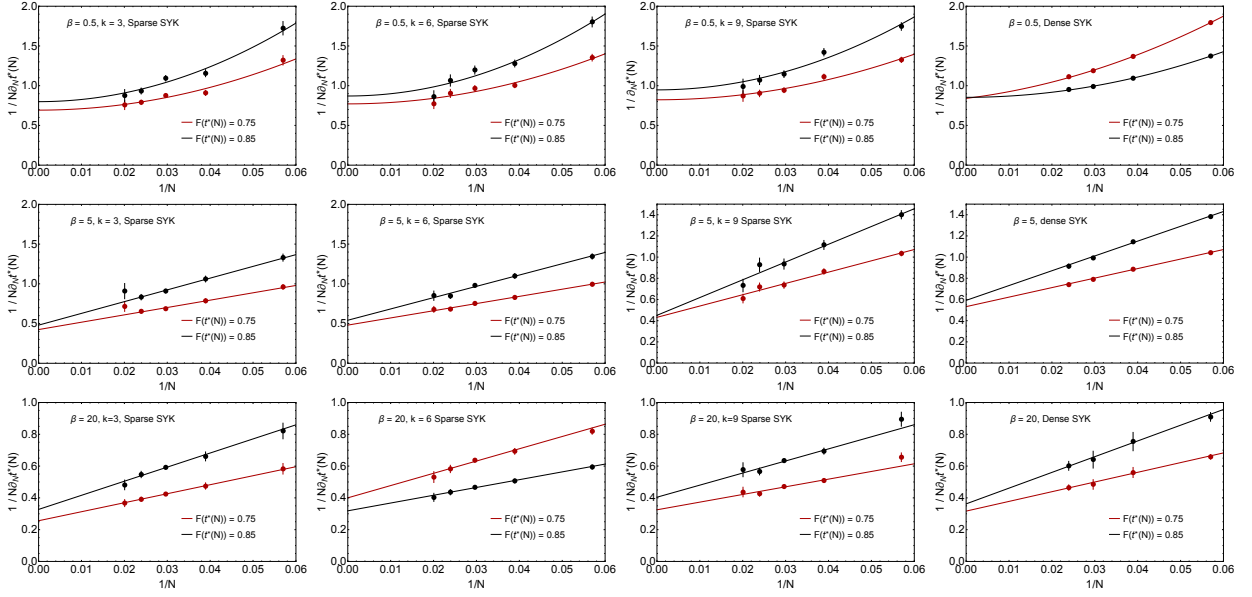

FIG. S1. The inverse slope $1/(N\partial_N t^*(N))$ versus $1/N$ for $F(t^*(N)) = 0.75$ (red) and $F(t^*(N)) = 0.85$ (black). Results are given for $\beta = 0.5$ (upper row), $\beta = 5$ (middle row) and $\beta = 20$ (lower row). The Lyapunov exponent is given by the extrapolation to $1/N \to 0$. The results for the two choices of $F_0$ seem to converge.

We observe that the computation time has a large dependence on $t_{\max}$, since for larger evolution time the Krylov algorithm takes substantially longer to converge. Additionally, for sparse SYK, the computation time has a large fluctuation, which can sometimes be as high as 30% of the average computation time. The fluctuations for dense SYKs are much lower, but nonetheless exist.

## II. DETAILS OF THE FINITE SIZE SCALING ANALYSIS

We now address two aspects of the numerical calculation of the Lyapunov exponent that are not included in the main text: the finite $N$ scaling analysis and the functional form of the OTOC to perform this analysis. .

### A. Calculation of the Lyapunov exponent from the slope of $t^*(N)$

The Lyapunov exponent can be obtained from the slope of $t^*(N)$ versus $\log N$, but for the range of times and sizes we can reach numerically, as mentioned in the main text, the Lyapunov exponent term has a residual $N$-dependence. Ideally, we fit observables for which this residual $N$-dependence is minimized. In agreement with [S1], the numerical results show, see Fig. S1, that for $\beta \geq 5$ the $1/N$ dependence of the inverse slope is close to linear so that the

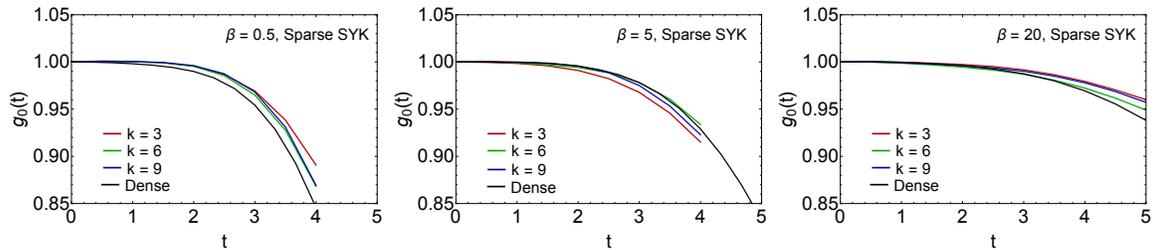

FIG. S2. The thermodynamic limit of the OTOC Eq. (2) obtained by linear extrapolation in $1/N$. We observe a residual time dependence, but only a weak dependence on $k$.

Lyapunov exponent is determined by

$$\frac{1}{Ndt^*(N)/dN} = \lambda_L + \frac{\alpha_1}{N} + +\frac{\alpha_2}{N^2} + O(1/N^2). \tag{S2}$$

For $\beta = 5$ and $\beta = 20$, an estimate for the Lyapunov exponent is given by the extrapolation of $1/(Ndt^*(N)/dN)$ to $1/N \to 0$ (in these cases we put $\alpha_2 = 0$). For small $\beta$, as also observed in [S1], the inverse slope depends quadratically on $1/N$, and to reduce the number of fitting parameters, we put $\alpha_1 = 0$ (except in the dense case where the accuracy of the data is sufficient to perform a 3-parameter fit) (see last column of Fig. S1).

Because of the much larger fluctuations in the sparse case [S2] we determine the slope $\partial_N t^*(N)$ from the difference of $t^*(N + 8)$ and $t^*(N)$ instead of using a discrete derivative involving $t^*(N)$ and $t^*(N - 2)$ as was done in [S1]. In Fig. S1 we show the results for $\beta = 0.5$, 5, 20 and values of $k$ as in the legend of the figure. We observe that the results for $F_0 = 0.75$ and $F_0 = 0.85$ seem to converge to the same point. The difference between the two is a measure of the systematic error of the procedure.

### B. Time independence of $g_0(t)$

In this section, we justify numerically the approximation of considering a time independent $g_0(t)$ in the calculation of the Lyapunov exponent. The exponential part of the OTOC is of order $1/N$ and vanishes in the large $N$ limit at fixed $t$. The function $g_0(t)$ is therefore given by

$$g_0(t) = \lim_{N \to \infty} F_N(t). \tag{S3}$$

Numerically, it is determined by a linear extrapolation in $1/N$. The results for $\beta = 0.5$, $\beta = 5$ and $\beta = 20$ are shown in Fig. S2. To satisfy the requirement $0.95 \lessapprox g_0(t) < 1$ we need that $t^*(N) \lessapprox 3$ for $\beta = 0.5$, $t^*(N) \lessapprox 4$ for $\beta = 5$ and $t^*(N) \lessapprox 5$ for $\beta = 20$. This is the case if we choose $F_0 > 0.75$. See Fig. 2.

To leading order in $1/N$ we thus have that

$$N\delta F(t) = N(F_N(t) - g_0(t)) \tag{S4}$$

does not depend on $N$. In Fig. S3 we show $\delta F$ as a function of $t$ for $N = 22$, 30, 38, 46 and 54 and $\beta = 0.5$, 5 and 20 both for $k = 6$ (left) and for the dense case (right). For $\beta \geq 5$ the curves are close, which justifies the neglect of $1/N^2$ corrections to the OTOC. Ideally, we would like use $\delta F$ to determine the Lyapunov exponent, but the extrapolation procedure results in large errors that make the determination of the Lyapunov exponent inaccurate. In addition, $\delta F$ has only a small concave region that can be approximated by an exponential function at all. Altogether, we did not find a way to reliably determine the Lyapunov exponent from $\delta F$. That is why we have adopted the method introduced in [S1].

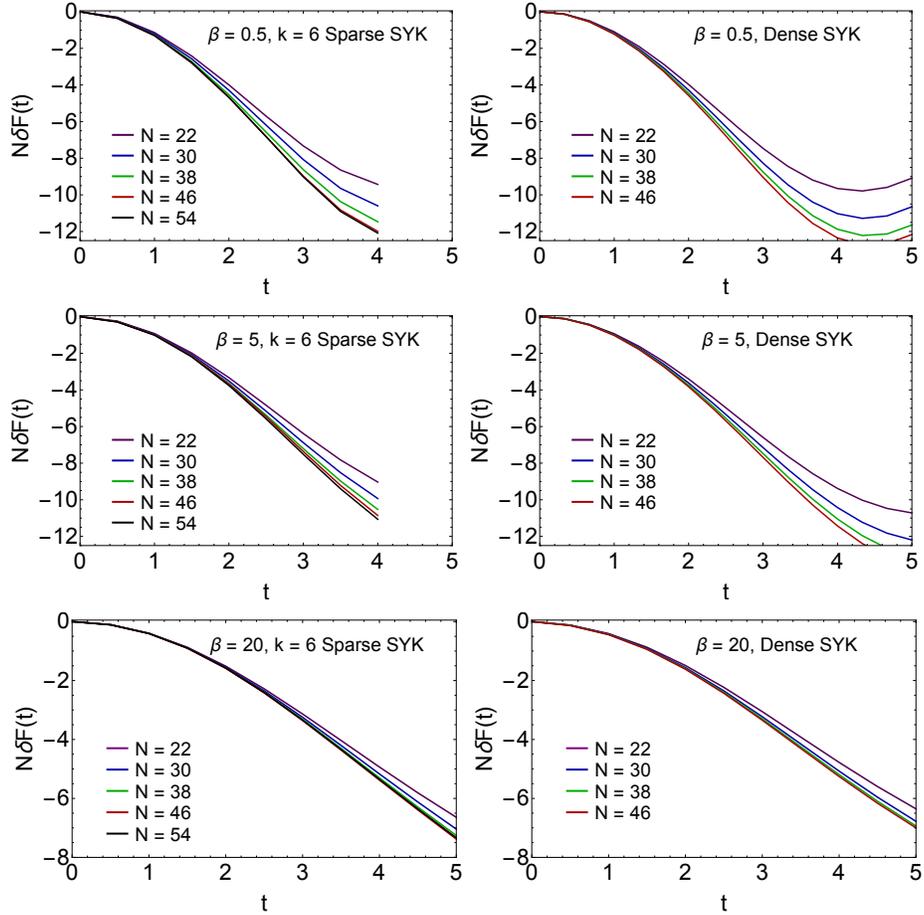

FIG. S3. Plots of the $N$ times difference $\delta F(t)$ of the OTOC and its asymptotic value $g_0(t)$, see Fig. S2. Only for small $\beta$ do we observe an $N$ dependence indicating the presence of $1/N^2$ corrections.


\fi

\end{document}